\documentclass[letter]{aa}  

\usepackage{graphicx}
\usepackage{txfonts}
\usepackage[version=4]{mhchem}
\usepackage{comment}
\usepackage{eqnarray}
\usepackage{placeins}
\usepackage{booktabs}
\usepackage{latexsym}
\usepackage{dcolumn}
\usepackage{amsmath}
\usepackage{mathtools}
\usepackage{tablefootnote}

\begin{document}

   \title{Detection of Interstellar \ce{H2CCCHC3N}}

   \subtitle{A Link Between Chains and Rings in Cold Cores?}

   \author{
        C. N. Shingledecker\inst{1,2,3}
        \and
        K. L. K. Lee\inst{4}
        \and
        J. T. Wandishin\inst{1}
        \and
        N. Balucani\inst{5}
        \and
        A. M. Burkhardt\inst{6}
        \and
        S. B. Charnley\inst{7}
        \and
        R. Loomis\inst{8}
        \and
        M. Schreffler\inst{1}
        \and
        M. Siebert\inst{9}
        \and
        M. C. McCarthy\inst{6}
        \and
        B. A. McGuire\inst{4,8,9}
          }

   \institute{Department of Physics and Astronomy, Benedictine College, Atchison, KS 66002, USA\\
              \email{cshingledecker@benedictine.edu}
        \and
            Zentrum für astrochemische Studien, Max Planck-Institut für extraterrestrische Physik, Garching bei München, Germany
        \and
            Institute for Theoretical Chemistry, Universität Stuttgart, Stuttgart, Germany
        \and
             Department of Chemistry, Massachusetts Institute of Technology, Cambridge, MA 02139, USA\\
             \email{bmcguire@mit.edu}
        \and
            Dipartimento di Chimica, Biologia e Biotecnologie, Università degli Studi di Perugia, 06123 Perugia, Italy
        \and
             Center for Astrophysics $\mid$ Harvard~\&~Smithsonian, Cambridge, MA 02138, USA
        \and
            Astrochemistry Laboratory and the Goddard Center for Astrobiology, NASA Goddard Space Flight Center, Greenbelt, MD 20771, USA
        \and
             National Radio Astronomy Observatory, Charlottesville, VA 22903, USA
        \and
            Department of Astronomy, University of Virginia, Charlottesville, VA 22904, USA
             }

   \date{Received ---; accepted ---}

 
  \abstract
   {The chemical pathways linking the small organic molecules commonly observed in molecular clouds to the large, complex, polycyclic species long-suspected to be carriers of the ubiquitous unidentified infrared emission bands remain unclear.  }
   {To investigate whether the formation of mono- and poly-cyclic molecules observed in cold cores could form via the bottom-up reaction of ubiquitous carbon-chain species with, e.g. atomic hydrogen, a search is made for possible intermediates in data taken as part of the GOTHAM (GBT Observations of TMC-1 Hunting for Aromatic Molecules) project.}
   {Markov-Chain Monte Carlo (MCMC) Source Models were run to obtain column densities and excitation temperatures. Astrochemical models were run to examine possible formation routes, including a novel grain-surface pathway involving the hydrogenation of C$_6$N and HC$_6$N, as well as purely gas-phase reactions between C$_3$N and both propyne (CH$_3$CCH) and allene (CH$_2$CCH$_2$), as well as via the reaction \ce{CN + H2CCCHCCH}.}
   {We report the first detection of cyanoacetyleneallene (\ce{H2CCCHC3N}) in space toward the TMC-1 cold cloud using the Robert C. Byrd 100 m Green Bank Telescope (GBT). Cyanoacetyleneallene may represent an intermediate between less-saturated carbon-chains, such as the cyanopolyynes, that are characteristic of cold cores and the more recently-discovered cyclic species like cyanocyclopentadiene.  Results from our models show that the gas-phase allene-based formation route in particular produces abundances of H$_2$CCCHC$_3$N that match the column density of $2\times10^{11}$ cm$^{-2}$ obtained from the MCMC Source Model, and that the grain-surface route yields large abundances on ices that could potentially be important as precursors for cyclic molecules.}
   {}


   \maketitle
%

\section{Introduction}

Highly unsaturated linear carbon chain molecules - such as the cyanopolyynes \ce{HC3N} \citep{turner_detection_1971}, \ce{HC5N} \citep{avery_detection_1976,broten_evidence_1976}, \ce{HC7N} \citep{kroto_detection_1978,little_observations_1978,winnewisser_detection_1978}, \ce{HC9N} \citep{broten_detection_1978}, and most recently, \ce{HC11N} \citep{Loomis:2021aa} - have long been known components of interstellar clouds and characteristic products of the unique gas-phase chemistry that occurs in those regions. A much more recent development has been the definitive detection of cyclic, aromatic molecules - such as benzonitrile \citep{mcguire_detection_2018} and cyanocyclopentadiene \citep{mccarthy_exhaustive_2020} - in even cold cores such as TMC-1. The presence of these kinds of cyclic, aromatic molecules is intriguing, in part, because it is as yet unclear whether they form via either a top-down mechanism from larger polycyclic species formed in circumstellar envelopes, or via bottom-up pathways from small precursors. 

If interstellar cyclic and polycyclic species can indeed form via a bottom-up route, then there may exist a direct chemical link to carbon-chain molecules such as the cyanopolyynes. In a previous work \citep{loomis_non-detection_2016}, we noted that, for very unsaturated linear molecules with more than $\sim10-11$ carbon atoms, the cyclic form can be more energetically stable, and thus, that such long carbon-chains might naturally tend to form rings at or above that size. Here, we envision a somewhat different mechanism in which shorter carbon chains (in this case consisting of a six-carbon-atom backbone) could hypothetically achieve aromaticity through a series of exothermic H-additions on grain surfaces, which would gradually break the linearity of the molecule and allow for eventual cyclization.

If grain-surface H-addition to highly unsaturated molecules such as cyanopolyynes were indeed a mechanism for forming aromatic species, then in principle, one would expect a fraction of the intermediate association products to desorb into the gas phase, where they could be observed. Such intermediates would be more highly saturated than, e.g. the cyanopolyynes but possess fewer H-atoms than the (poly)cyclic form.

In this work, we report the first detection of cyanoacetyleneallene, \ce{H2C=C=CHC#CC#N}, (IUPAC name 4,5-hexadien-2-ynenitrile, SMILES \ce{C=C=CC\#CC\#N}) in the cold core TMC-1. This species, recently analyzed in the laboratory by \citet{mccarthy_exhaustive_2020}, could plausibly represent such a chemical ``missing link.'' The first detection of the unsubstituted form of this molecule, allenyl acetylene (\ce{H2CCCHCCH}), also in TMC-1, has also recently been reported by \citet{cernicharo_discovery_2021-2} - who have recently detected a number of new interstellar species in that source using data taken with the Yebes 40 m radiotelescope \citep{cernicharo_discovery_2021-1,cernicharo_discovery_2021,marcelino_study_2021,cabezas_space_2021,agundez_discovery_2021}, thereby allowing for further examinations of their abundances and potential chemical connections. 

The rest of this paper is organized as follows: \S\ref{sec:observations} describes our observations, in \S\ref{sec:analysis} we summarize our method of analysis, an overview of our astrochemical modeling is given in \S\ref{sec:modeling}, and finally, in \S\ref{sec:conclusions} we present the main conclusions of this work.

\begin{figure*}[ht!]
    \centering
    \includegraphics[width=\textwidth]{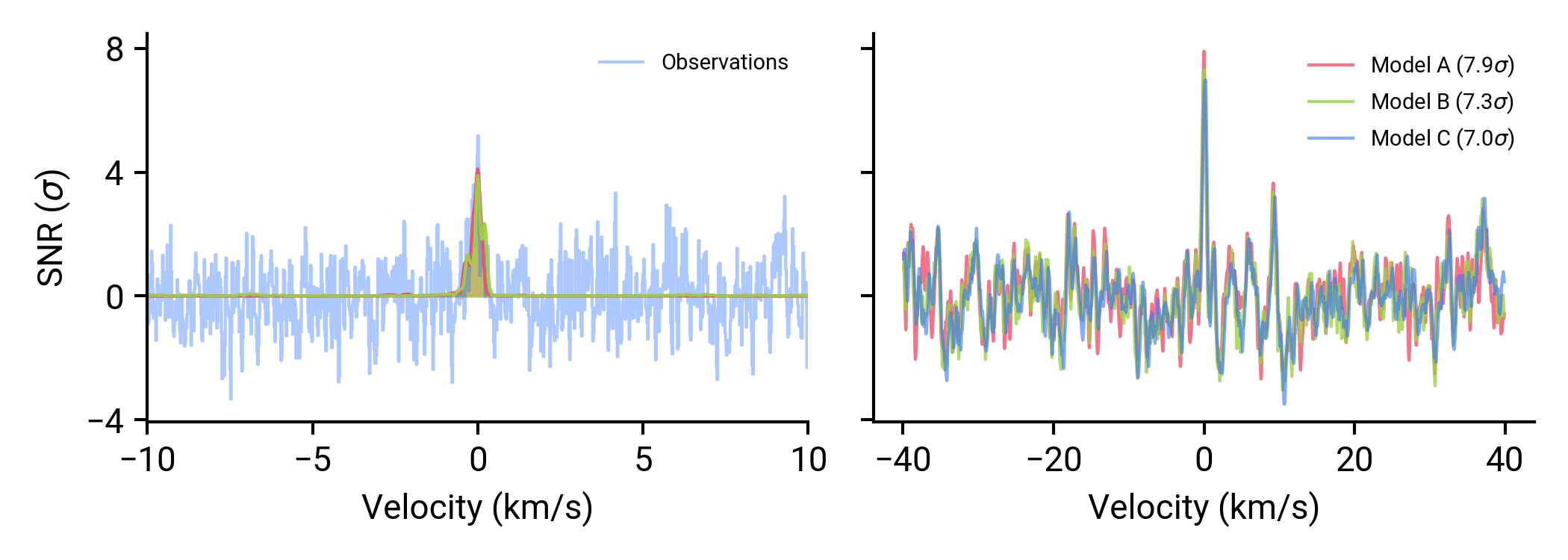}
    \caption{Left: velocity stack of the observations (blue) and the two MCMC models. Right: matched filter responses of three models considered; see text for details.}
    \label{fig:matched_filter}
\end{figure*}

\section{Observations} \label{sec:observations}

Details of the observing strategy for these observations have been presented elsewhere \citep{McGuire:2020bb,mcguire_detection_2018} and will only be summarized here.  The spectral line survey data analyzed here were taken as part of the GOTHAM (GBT Observations of TMC-1: Hunting Aromatic Molecules) project using the 100 m Robert C. Byrd Green Bank Telescope (GBT; \citealt{McGuire:2020bb}).  The full dataset used includes observations from GBT projects GBT17A-164, GBT17A-434, GBT18A-333, GBT18B-007, and data from project GBT19A-047 acquired through 2020 June. The pointing center for the observations was the cyanopolyyne peak of TMC-1 at (J2000) $\alpha$~=~04$^h$41$^m$42.50$^s$ $\delta$~=~+25$^{\circ}$41$^{\prime}$26.8$^{\prime\prime}$.  Pointing accuracy was confirmed every 1--2 hours using the nearby calibrator source J0530+1331; typical pointing convergence was $\lesssim$5$^{\prime\prime}$.  All observations were taken in position-switched mode, with a 1$^\circ$ throw to an off position verified to contain no emission. The spectra were flux calibrated using an internal noise diode and through complementary VLA observations of the flux-calibrator source J0530+1331 \citep{McGuire:2020bb}; the flux uncertainty is expected to be $\sim$20\%.

\section{Analysis Techniques and Results} \label{sec:analysis}

Details of the data analysis techniques, specifically the procedures for spectral stacking and matched filtering, are presented in \citet{Loomis:2021aa} and we only briefly discuss the relevant parts here. With a catalog of precise rest frequencies for cyanoacetyleneallene, we performed an initial search for the strongest transitions from a nominal 8 K simulation within the coverage of our DR2 data, with no individual transitions observable above an appreciable signal-to-noise ratio (SNR; 5$\sigma$). We subsequently performed a velocity stack, whereby we calculate a noise- and intensity-weighted average based on chunks of spectra centered at each catalog frequency, combining the flux spread across many thousands of spectral channels. 

To improve the SNR of this velocity stack, we then carry out an independent forward modeling of the molecular flux using Markov Chain Monte Carlo (MCMC)---using a set of physical parameters based on prior work, this comprises the column density ($N_\mathrm{col}$, cm$^{-2}$), source size, and radial velocity (VLSR, km/s) for four velocity components \citep{Dobashi:2018kd,Dobashi:2019ev,McGuire:2020bb}, and a shared excitation temperature ($T_\mathrm{ex}$, K) and Gaussian linewidth ($dv$, km/s). The source size is used to calculate a beam dilution factor that modifies the simulated flux based on how well the spatial extent of the source matches with the GBT beam. For computational efficiency, we select transitions from the catalog within 1\% of the largest predicted intensity: this corresponds to a total of 857 transitions used for our analysis, spanning the frequency range of our observational data. The procedures described are built into the \textsc{molsim} package \citep{lee_molsim_2020}, which performs the spectral simulation and wraps the affine-invariant MCMC implementation of \textsc{emcee} \citep{foreman-mackey_emcee_2013} and posterior analysis routines from \textsc{ArviZ} \citep{kumar_arviz_2019}.

The MCMC sampling is subsequently used to condition the simulations by evaluating the likelihood of any given set of parameters (which constitute a model), guided by a set of prior distributions. As the choice of prior is often a topic of intense scrutiny in Bayesian approaches, we explored the choice of prior as part of our analysis. For molecules with individual observable transitions, we use weakly informative uniform priors to guide sampling: the resulting posterior is bootstrapped and used as a prior for MCMC sampling of molecules \emph{without} observable transitions. The choice of which molecule to use as a ``template'' is typically motivated by chemical similarity, assuming they have similar chemical origins (i.e. shared pathways) and physical conditions. 

To assess the sensitivity of our models to the choice of prior, we performed the MCMC sampling using \ce{HC9N} and benzonitrile posteriors as prior distributions for cyanoacetyleneallene; we refer to the posterior means of these cases as Source Models A and B for later discussion. From our prior work \citep{Loomis:2021aa}, we have found that the column densities are often highly covariant with the source size due to beam dilution; in this work, we leave the source sizes fixed to their respective prior mean values as a means to improve sampling convergence, to a total of ten free parameters. The simulated spectra are subsequently velocity stacked and cross-correlated with the observational velocity stack to determine a matched filter response, representing a lower bound on the significance of our detection. As with other GOTHAM detections, we adopt a $5\sigma$ cutoff for the significance criterion to claim a detection \citep{Loomis:2021aa, lee_discovery_2021}. Finally, to ensure that the MCMC models are not artificially overestimating the matched filter response, we tested a third case, Source Model C, whereby we used the posterior parameters of benzonitrile to perform the spectral simulation \emph{without} further sampling or optimization. These results are shown in Figure \ref{fig:matched_filter}, and a more complete set of MCMC results including corner plots can be found in the Appendix.

The left panel of Figure \ref{fig:matched_filter} shows the velocity stack of the observations, and of the simulations using the posterior means of Source Models A and B: both models are able to reproduce the observed spectra faithfully albeit with very different posterior parameters. Source Model A, which uses \ce{HC9N} as its prior, corresponds to a larger derived column density (combined 9$\times$10$^{13}$\,cm$^{-2}$) and an extremely low excitation temperature of 2.8\,K, close to the cosmic microwave background temperature \citep{fixsen_temperature_2009}, while Source Model B corresponds to a more typical column density (combined $2\times10^{11}$\,cm$^{-2}$) and excitation temperature (8.5\,K), compared to similar molecules recently detected toward TMC-1 \citep{lee_discovery_2021}.

On the right panel of Figure \ref{fig:matched_filter} are the matched filter spectra, which are the cross-correlation of the observational and simulated velocity stacks. The peak impulse response can be interpreted as a lower bound to the detection significance, although given that each model produces a similar impulse response, we are unable to rely on this metric for model selection. Despite being physically unintuitive, Source Model A results in the largest peak impulse response, while Source Model B only marginally improves the SNR from the baseline Source Model C, despite being conditioned on the observations. The problem of which model to base assumptions on is one of model selection, and would require a means to calculate the full Bayesian evidence by integration over model space: this is not readily tractable given the complexity of our model. Thus, while we are unable to definitively provide estimates of the modeling parameters, the significant impulse response attained even by the unoptimized, fiducial Source Model C provides strong evidence for the velocity stacked detection of cyanoacetyleneallene toward TMC-1 at $\geq 7\sigma$.

While we are unable to quantitatively rule out each Source Model with the matched filter response, we can provide qualitative arguments based on similarity to other detected molecules in TMC-1, particularly in terms of the derived column density. Thus far, molecules comprised of five to seven heavy atoms, e.g. the [\ce{H3C5N}] family \citep{lee_discovery_2021} and c-\ce{C5H5CN} \citep{lee_discovery_2021,McCarthy:2021aa}, have all been detected with column densities of approximately $\sim10^{11}$\,cm$^{-2}$. The molecule we report the detection of in this work, cyanoacetyleneallene corresponds to the \ce{-CN} derivative of a recently detected molecule, allenyl acetylene, by \citet{cernicharo_discovery_2021}: if we assume these molecules are related via the straightforward radical addition of \ce{CN}, as shown in reaction (R8) in \S4, then we can compare their column density ratios with other hydrocarbon/\ce{CN} pairs with similar bond structure detected in TMC-1. Because the allenyl acetylene was detected with observations at higher frequencies (34--50\,GHz) compared to the GOTHAM survey, direct abundance comparisons should be taken with a degree of caution as they can be highly degenerate with the assumed or fitted source size.  \citet{cernicharo_discovery_2021-2} assumed a source size radius of 40$^{\prime\prime}$ for allenyl acetylene, while the derived FWHM source sizes of the individual velocity components for cyanoacetyleneallene here ranged from 70--259$^{\prime\prime}$ for Model B. Nevertheless, as shown in Table \ref{tab:cnratios}, the hydrocarbon to \ce{CN} derivative ratio ranges from 30--120, with a large variation due to a combination of different telescopes, analysis techniques, and production routes---it is clear, however, that the column density derived from Source Model B is most consistent with previous determinations, whereas Source Model A represents a scenario not yet seen toward this source. Given, however, that we are not aware of any obvious reasons for such a significant deviation in chemical trends, we believe it is more likely than not that Source Model B is a better approximation to the chemical physics of cyanoacetyleneallene in TMC-1. This relatively consistent R-H/R-CN ratio has been discussed previously for the methyl-terminated carbon chains \citep{remijan_methyltriacetylene_2006}. This analysis provides a strong motivation to further study the chemical relation between R-H and R-CN pairs, for which methyl-terminated carbon chains could provide an abundant family of molecules to study this relation in detail. Futhermore, since the radical addition of \ce{CN} has been found to be efficient in interstellar conditions \citep{huang_crossed_1999,balucani_crossed_1999,balucani_combined_2009,balucani_laboratory_2000,kaiser_formation_2001,Lee:2019dh,Cooke:2020we}, this relation could be crucial for understanding the relation between pure hydrocarbon rings and their recently detected cyano-derivatives \citep{mcguire_detection_2018,McCarthy:2021aa,Lee:2021bb,Burkhardt:2021aa,McGuire:2021aa}.

\begin{table*}[tb]
    \centering
    \caption{Column density ratios of pure hydrocarbon and their \ce{CN}-derivatives detected towards TMC-1.}
    \begin{tabular}{r@{/}l c r}
    \toprule
    \multicolumn{2}{c}{Molecular pair} & R-H/R-CN ratio & Reference  \\
    \midrule 
    \ce{H2CCCHCCH}&\ce{H2CCCHC3N} & 0.1$^a$, 60$^b$ & \citet{cernicharo_discovery_2021}; this work \\
    \ce{CH3C2H}&\ce{CH2(CN)CCH} & ${\sim}120$ & \citet{snyder_confirmation_2006,McGuire:2020bb} \\
    \ce{CH3C2H}&\ce{CH3C3N} & ${\sim}64$ & \citet{snyder_confirmation_2006,remijan_methyltriacetylene_2006} \\
    \ce{CH3C4H}&\ce{CH3C5N} & ${\sim}35$ & \citet{snyder_confirmation_2006,remijan_methyltriacetylene_2006} \\
    \ce{CH2CCCH}&\ce{CH2CCCCN} & ${\sim}100$ & \citet{cernicharo_discovery_2021a,lee_discovery_2021} \\
    \ce{CH2CCCH}&\ce{HCCCHCHCN} & ${\sim}50$ & \citet{cernicharo_discovery_2021a,lee_discovery_2021} \\
    \bottomrule
    \end{tabular}
    \\
    $^a$ Based on Source Model A, with a column density of $9\times10^{13}$\,cm$^{-2}$\\
    $^b$ Based on Source Model B, with a column density of $2\times10^{11}$\,cm$^{-2}$
    \label{tab:cnratios}
\end{table*}

\section{Astrochemical Modeling} \label{sec:modeling}

Astrochemical models were used to explore possible formation routes that could explain the observed abundance of cyanoacetyleneallene. For this, we employed the three-phase (gas, grain/ice surface, and ice bulk) \texttt{NAUTILUS} v1.1 code \citet{ruaud_gas_2016}. The chemical network and physical conditions we use are based on those described in other recent GOTHAM project papers \citep{Loomis:2021aa,McGuire:2020bb}. Briefly, we ran 0D simulations using constant physical conditions relevant to TMC-1, including a gas/dust temperature of 10 K, standard cosmic ray ionization rate of $\zeta=1.3\times10^{-17}$ s$^{-1}$, and standard chemical desorption efficientcy of 1\%. Initial elemental abundances are generally taken from \citet{hincelin_oxygen_2011}, with the exception of the initial C/O ratio, which we assume to be $\sim 1.1$, which was found to yield best agreement between our models and the abundances of the cyanopolyynes \citep{Loomis:2021aa}. 

To the GOTHAM network, we have added a number of gas- and grain-reactions related to cyanoacetyleneallene. As a preliminary test of whether the cyclic molecules detected in TMC-1 could have formed via grain-surface H-addition reactions to unsaturated carbon-chain species -- and whether cyanoacetyleneallene might be an intermediate in this process -- we have included the following reaction pathway:

\begin{equation}
    \ce{C6N + H -> HC6N}
    \label{r1}
    \tag{R1}
\end{equation}

\begin{equation}
    \ce{HC6N + H -> H2C6N}
    \label{r2}
    \tag{R2}
\end{equation}

\begin{equation}
    \ce{HC6N + H -> HCCCHC3N}
    \label{r3}
    \tag{R3}
\end{equation}

\begin{equation}
    \ce{H2C6N + H -> H2CCCHC3N}
    \label{r4}
    \tag{R4}
\end{equation}

\begin{equation}
    \ce{HCCCHC3N + H -> H2CCCHC3N}
    \label{r5}
    \tag{R5}
\end{equation}

\noindent
where here, the grain-surface H-addition reactions are assumed to proceed barrierlessly. There are a number of possible additional products that have not been included. Since our goal in this work is to estimate the feasibility of such processes as a viable formation route to (ultimately) cyclic molecules, we have simplified the chemistry to just the above-mentioned reactions. Thus, abundances of \ce{H2CCCHC3N} formed via \eqref{r1} - \eqref{r5} will likely be somewhat lower than what we predict in our simulations.

In addition to the grain surface route given in (R1) - (R5), we have also included three relevant gas-phase formation routes, given by (R6b), (R7e), and (R8). Two of the likely candidates, e.g. reactions between \ce{C3N} and \ce{CH2CCH2} and \ce{CH3CCH}, were studied by \citet{fournier_reactivity_2014} who measured the respective rate coefficients for temperatures in the range of 24 to 300 K. Fits of their experimental data to the Arrhenius-Kooij formula, given by

\begin{equation}
k = \alpha \ \left( \frac{T}{300\,\mathrm{K}} \right)^{\beta} \ \exp \left( \frac{\gamma}{T} \right) \ \textrm{cm}^3 \textrm{molecule}^{-1} \textrm{s}^{-1}.
\label{kr7}
\end{equation}

   \begin{table}
      \caption[]{Arrhenius-Kooij Parameters for the reaction between \ce{C3N} and hydrocarbons from \citet{fournier_reactivity_2014}.}
         \label{tab:fournier}
     $$ 
         \begin{array}{p{0.4\linewidth}ccc}
            \hline
            \noalign{\smallskip}
            Reaction      &  \alpha & \beta & \gamma \\
                          &  (\mathrm{cm}^{3}\;\mathrm{s}^{-1}) &      & (\mathrm{K}) \\
            \noalign{\smallskip}
            \hline
            \noalign{\smallskip}
            \ce{C3N + CH2CCH2} & 5.243\times10^{-10} & -0.575 & -50.53  \\
            \ce{C3N + CH3CCH} & 4.617\times10^{-10} & -0.407 & -32.42  \\
            \noalign{\smallskip}
            \hline
         \end{array}
     $$ 
   \end{table}

\noindent
lead to the parameters shown in Table \ref{tab:fournier}. From the values shown in Table \ref{tab:fournier}, one can see that reactions (R6) and (R7) are very efficient at even the low temperatures of cold cores. Unfortunately, \citet{fournier_reactivity_2014} were not able to establish the nature of the products and their branching ratios based on their data from the CRESU experiments, but suggested (R6a)-(R6b) and (R7a)-(R7d) as channels, given below.

\begin{equation*}
\begin{array}{lllrr}
       \ce{C3N + CH2CCH2} & \hspace{-1em} \ce{-> HC3N + HCCCH2} & & & \mathrm{(R6a)} \\
                         &  \hspace{-1em} \ce{-> H + H2CCCHC3N} & & & \mathrm{(R6b)} \\
\end{array}
\end{equation*}

\begin{equation*}
\begin{array}{lllrr}
       \ce{C3N + CH3CCH} & \hspace{-1em} \ce{-> HC3N + C2CH3} & & & \mathrm{(R7a)} \\
                         &  \hspace{-1em} \ce{-> H3C6N + H} & & & \mathrm{(R7b)} \\
                         &  \hspace{-1em} \ce{-> HC3N + HCCCH2} & & & \mathrm{(R7c)} \\
                         &  \hspace{-1em} \ce{-> HCCCH2C3N + H} & & & \mathrm{(R7d)} \\
                         &  \hspace{-1em} \ce{-> H + H2CCCHC3N} & & & \mathrm{(R7e)} \\
\end{array}
\end{equation*}

Though Fournier and coworkers did not include \ce{H2CCCHC3N} as a possible product of (R7), by analogy to similar reactions involving CN, we have assumed that formation of cyanoacetyleneallene (R7e) is in fact also a possible product channel.
Here, we have assumed equal branching fractions among all product channels based on analogous reactions and previous crossed beam studies by \citet{huang_crossed_1999}.

Finally, given the recent detection of acetylene allene by \citet{cernicharo_discovery_2021-2}, we have included the following third gas-phase formation route to cyanoacetyleneallene 

\begin{equation}
    \ce{CN + H2CCCHCCH -> H2CCCHC3N + H.}
    \label{r8}
    \tag{R8}
\end{equation}

\noindent
Here, we have assumed a rate coefficient of $k_\mathrm{R8}=(1/3)\times k_\mathrm{R6}$, since (R8) and (R6) involve the same number of atoms and share a similar reaction mechanism - implying that (R8) likewise has no entrance barrier and very similar long range interactions to (R6). The $1/3$ factor accounts for two other possible competing product channels. We note that we have disregarded the reaction of the propargyl radical (C$_3$H$_3$) with cyanoacetylene based on results by \citet{da_silva_mystery_2017}, who found evidence of an entrance barrier in the reaction C$_3$H$_3$+C$_2$H$_2$.

\vspace{1em}

In addition to the previously noted formation routes, we have also included destruction reactions with the ions \ce{H3+}, \ce{He+}, \ce{HCO+}, and \ce{H3O+}, where the Langevin formula was used to estimate rate coefficients \citep{woon_quantum_2009}. Gas-phase cyanoacetyleneallene is also depleted via accretion onto grains.

   \begin{figure}
   \centering
   \includegraphics[width=\hsize]{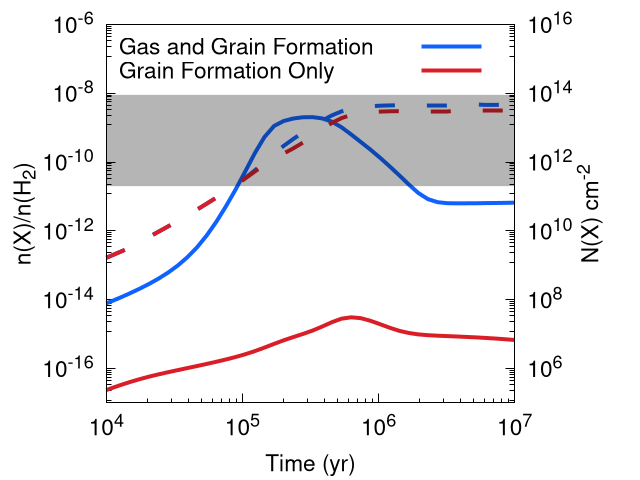}
    \caption{Abundance of \ce{H2CCCHC3N} both in the gas (solid lines) and in the mantle (dashed lines). Blue lines correspond to the model in which \ce{H2CCCHC3N} was formed both in the gas and on grains, while red lines show results from the model in which cyanoacetyleneallene was produced only on grains. The range of column densities derived from Source Models A \& B are represented by the horizontal gray bar.}
    \label{fig:modeling}
   \end{figure}

The results of our simulations are shown in Fig. \ref{fig:modeling}. There, we present results from two models. In the first, represented by blue lines, \ce{H2CCCHC3N} was formed via both the grain-surface and gas-phase reactions. In the second, represented by red lines, \ce{H2CCCHC3N} was made solely via the grain-surface route. 

A comparison between the calculated abundances of cyanoacetyleneallene from the two models shows that inclusion of the gas-phase formation routes substantially increases its abundance over the predictions from the grain-surface-only by up to seven orders of magnitude at model times of about $2\times 10^5$ yr. It should be noted that in the gas and grain formation model, reaction (R6b) dominates at all relevant simulation times. 

Given the range of derived column densities from the MCMC fits, comparison with astrochemical modeling results is not entirely straightforward. Nevertheless, with the gas-phase production routes included, the model is able to reproduce the more reasonable columns predicted by Source Model B, assuming $N(\ce{H2})=10^{22}$ cm$^{-2}$ \citep{ohishi_chemical_1998}, and even achieves a peak abundance within a factor of a few from the larger columns inferred from Source Model A. Taken together, our modeling data suggest that the gas-phase reactions are important formation routes for cyanoacetyleneallene.  

Returning to Fig. \ref{fig:modeling}, we note that in both models, the abundance of \ce{H2CCCHC3N} trapped in dust-grain ice mantles (dashed lines) is significant. A comparison between the two shows that, while the effect of gas-phase accretion of cyanoacetyleneallene does enhance the solid-phase abundance somewhat, most of the \ce{H2CCCHC3N} on grains is formed via the H-addition reactions given by \eqref{r1} - \eqref{r5}. Thus, while the gas-phase formation routes, and in particular (R6b), seem to be most important in explaining the observed gas-phase abundance of cyanoacetyleneallene in TMC-1, the grain-surface routes do seem to be efficient, and might yield precursors to cyclic molecules upon further surface reactions with atomic hydrogen. This provides a strong motivation to further develop a robust chemical network of grain surface saturation of carbon chains into cyclic molecules, which are currently underproduced in existing dark cloud chemical models \citep{Burkhardt:2021aa,McCarthy:2021aa,McGuire:2021aa}.

\section{Conclusions} \label{sec:conclusions}

In this work, we have reported the velocity-stacked detection of cyanoacetyleneallene (\ce{H2CCCHC3N}) in the cold core, TMC-1. Our MCMC modeling with priors from two representative species -- \ce{HC9N} and c-\ce{C6H5CN} -- yielded substantially different derived column densities, $9\times 10^{13}$ cm$^{-2}$ and $2\times10^{11}$ cm$^{-2}$, respectively. The disparate results are due to the strong covariance between column density and excitation temperature. In models using \ce{HC9N} as a prior, predicted excitation temperatures were found to be close to the CMB. Our MCMC model using c-\ce{C6H5CN} as a prior represents both a more physically reasonable case of small column densities and yielded an excitation temperature more typical of molecules in TMC-1. Moreover, the minimum matched filter response obtained was $7\sigma$ based on a fiducial model--- a significant and robust detection. Going forward, more informative priors placed on the sampling will significantly help constrain the model space.

In an attempt to investigate whether linear carbon-chain species such as \ce{C6N} and \ce{HC6N} could serve as precursors to cyclic molecules like cyanocyclopentadiene (c-\ce{C5H5CN}) -- and by extension and whether \ce{H2CCCHC3N} might represent an intermediate in this process -- we have included a number of grain-surface H-addition reactions leading to cyanoacetyleneallene. Results from our astrochemical models indicate that while this grain-surface route does lead to large amounts of \ce{H2CCCHC3N} on grain surfaces, the gas-phase reaction between \ce{C3N} and both propyne (\ce{CH3CCH}) and in particular allene (\ce{CH2CCH2}) are likely the dominant formation routes for the gas-phase cyanoacetyleneallene detected in TMC-1. 

Nevertheless, \ce{H2CCCHC3N} on grain surfaces could still react with H atoms, and therefore might yet represent an intermediate in a bottom-up formation route for the kinds of cyclic species now known in TMC-1 \citep{mcguire_detection_2018,McCarthy:2021aa,McGuire:2021aa}. Therefore, the preliminary grain-surface reactions presented here which connect the abundant carbon-chain species with cyclic molecules represent attractive targets for detailed study, either experimentally or via quantum-chemical calculations. 

Thus, our finding that c-\ce{C6H5CN}, rather than \ce{HC9N}, is a more physically realistic prior for our \ce{H2CCCHC3N} Source Models meshes nicely with our astrochemical modeling results since, taken together, they both seem to suggest that (a) long carbon-chain species like \ce{HC9N} are not direct precursors to the gas-phase cyanoacetyleneallene that was observed, but that (b) the grain-surface hydrogenation of cyanacetyleneallene is efficient and might still be a promising bottom-up pathway to cyclic species such as benzonitrile under cold-core conditions.

\begin{acknowledgements}
C.N.S. thanks V. Wakelam for use of the \texttt{Nautilus} v1.1 code. M.C.M and K.L.K.L. acknowledge financial support from NSF grants AST-1908576, AST-1615847, and NASA grant 80NSSC18K0396.  A.M.B. acknowledges support from the Smithsonian Institution as a Submillimeter Array (SMA) Fellow. S.B.C.was supported by the NASA Astrobiology Institute through the Goddard Center for Astrobiology.  The National Radio Astronomy Observatory is a facility of the National Science Foundation operated under cooperative agreement by Associated Universities, Inc.  The Green Bank Observatory is a facility of the National Science Foundation operated under cooperative agreement by Associated Universities, Inc. This research has made use of NASA's Astrophysics Data System.
\end{acknowledgements}

%
%
\bibliographystyle{aa}
\bibliography{references,brett_bib,kelvin_bib}

\begin{appendix} 

\renewcommand{\thefigure}{A\arabic{figure}}
\renewcommand{\thetable}{A\arabic{table}}
\renewcommand{\theequation}{A\arabic{equation}}
\setcounter{figure}{0}
\setcounter{table}{0}
\setcounter{equation}{0}

\section{MCMC posterior analysis}

\begin{figure*}
    \centering
    \includegraphics[width=\textwidth]{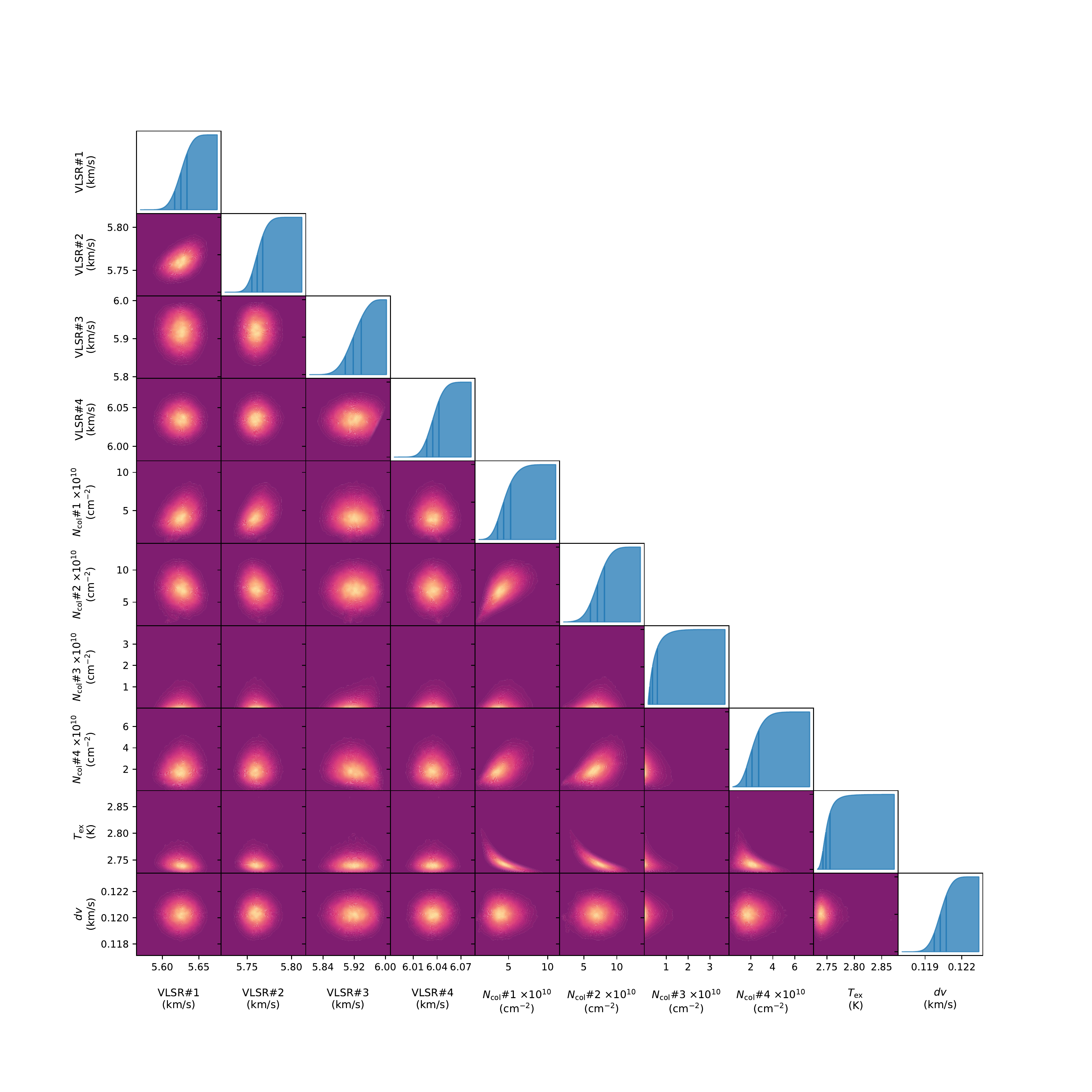}
    \caption{Corner plot for cyanoacetyleneallene using \ce{HC9N} as a prior, referred to as Model A in text. Traces on the diagonal correspond to cumulative density plots, which represent the integral over parameter space for the marginal likelihood of each parameter.}
    \label{fig:hc9n_prior}
\end{figure*}

\begin{figure*}
    \centering
    \includegraphics[width=\textwidth]{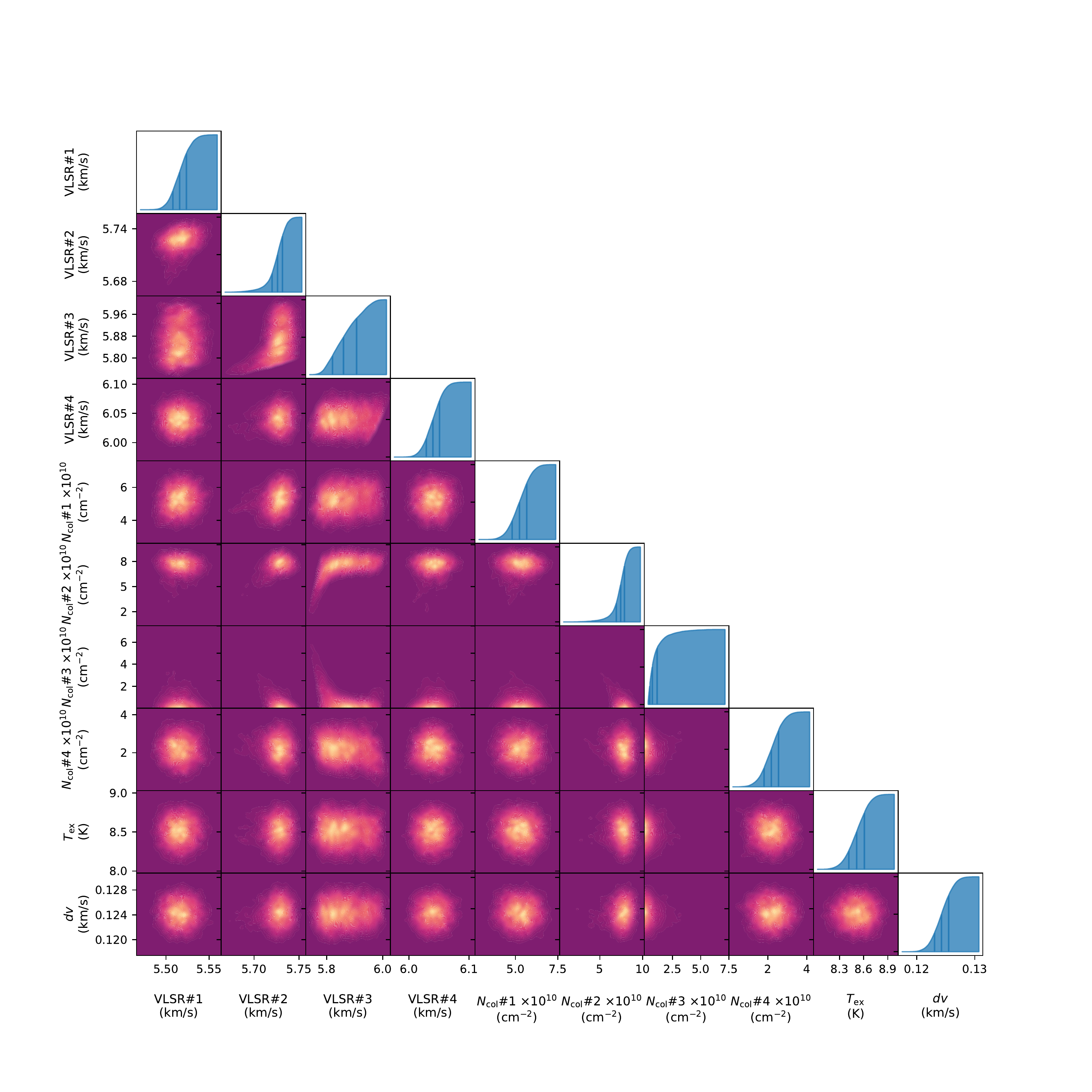}
    \caption{Corner plot for cyanoacetyleneallene using c-\ce{C6H5CN} as a prior, referred to as Model B in text. Traces on the diagonal correspond to cumulative density plots, which represent the integral over parameter space for the marginal likelihood of each parameter.}
    \label{fig:bn_prior}
\end{figure*}

Figures \ref{fig:hc9n_prior} and \ref{fig:bn_prior} show corner plots of the posterior distributions for cyanoacetyleneallene when using \ce{HC9N} (Model A) or benzonitrile (Model B) as informative priors. The diagonal traces correspond to empirical cumulative distribution function (ECDF) plots, which show the range of values considered in the MCMC sampling, while the off-diagonal contours reveal covariances between parameters. The source sizes are not shown in either plots, as they are fixed to their respective values and unchanged throughout the MCMC sampling.

An important aspect in Figure \ref{fig:hc9n_prior} is the extremely large covariance between excitation temperature ($T_\mathrm{ex}$) and column density ($N_\mathrm{col}$) for all four sources: higher excitation temperatures correspond to smaller column densities, and likely represents insufficient data coverage, which would help constrain the range of temperatures that would be consistent with the data. Table \ref{tab:mcmcparams} summarizes the MCMC results, and ultimately the parameters used for the spectral simulation and matched filtering.

\begin{table*}[h]
    \centering
    \caption{Summary statistics of posteriors values obtained from the MCMC sampling, as well as for the fiducial spectral simulation. Uncertainties correspond to the 2nd and 97th percentile posterior likelihood (i.e. the 95\% highest posterior densities); values without uncertainties are fixed to their prior values. Excitation temperatures and linewidths are shared amongst the four velocity components.}
    \begin{tabular}{c c c c}
    \toprule
    Velocity component & Model A\tablefootmark{a} & Model B\tablefootmark{b} & Model C\tablefootmark{c} \\
    \midrule
    \multicolumn{4}{c}{Source size ($\prime\prime$)} \\
    \midrule
     1 & 54 & 70 & 70 \\
     2 & 35 & 111 & 111 \\
     3 & 99 & 269 & 269 \\
     4 & 39 & 228 & 228 \\
    \midrule
    \multicolumn{4}{c}{Velocity ($v_{lsr}$---km/s)} \\ 
    \midrule
     1 & $5.620^{+0.026}_{-0.026}$ & $5.518^{+0.019}_{-0.018}$ & 5.575 \\
     2 & $5.762^{+0.019}_{-0.018}$ & $5.728^{+0.018}_{-0.018}$ & 5.767 \\
     3 & $5.918^{+0.060}_{-0.058}$ & $5.870^{+0.111}_{-0.097}$ & 5.892 \\
     4 & $6.035^{+0.023}_{-0.023}$ & $6.039^{+0.032}_{-0.033}$ & 6.018 \\
    \midrule
    \multicolumn{4}{c}{Column density ($N_\mathrm{col}$---cm$^{-2}$)} \\ 
    \midrule
     1 & $3.55^{+3.60}_{-2.63}\times10^{13}$ & $5.32^{+1.32}_{-1.34}\times10^{10}$ & $2.31 \times10^{11}$ \\
     2 & $4.81^{+2.42}_{-2.45}\times10^{13}$ & $7.66^{+1.69}_{-1.67}\times10^{10}$ & $6.00 \times10^{11}$ \\
     3 & $1.91^{+3.90}_{-1.91}\times10^{12}$ & $6.26^{+1.32}_{-6.26}\times10^{9}$ & $3.67 \times10^{11}$ \\
     4 & $1.24^{+1.24}_{-1.10}\times10^{13}$ & $2.16^{+1.13}_{-1.12}\times10^{10}$ & $5.33 \times10^{11}$ \\
    \midrule
    \multicolumn{4}{c}{Excitation temperature ($T_\mathrm{ex}$---K)} \\ 
    \midrule
    ~ & $2.8^{+0.0}_{-0.0}$ & $8.5^{+0.3}_{-0.3}$ & 8.9 \\
    \midrule
    \multicolumn{4}{c}{Linewidth ($dv$---km/s)} \\ 
    \midrule
    ~ & $0.1202^{+0.0014}_{-0.0013}$ & $0.1317^{+0.0172}_{-0.0167}$ & 0.125 \\
    \bottomrule
    \end{tabular}

    \tablefoottext{a}{Model A uses the posterior parameters of \ce{HC9N} as a prior---with the exception of source sizes, all other parameters are fit.}
    \tablefoottext{b}{Model B uses the posterior parameters of c-\ce{C6H5CN} as a prior---with the exception of source sizes, all other parameters are fit.}
    \tablefoottext{c}{Model C uses the posterior mean of c-\ce{C6H5CN} for the spectral simulation, however does no further optimization. Values adapted from \citet{McGuire:2021aa}. }
    \label{tab:mcmcparams}
\end{table*}

\end{appendix}

\end{document}